\documentclass[pre,preprint]{revtex4-1}
\usepackage{subfigure}
\usepackage[dvips]{color}
\usepackage{epsfig}
\usepackage{hyperref}
\usepackage{amsthm}
\usepackage{ae}

\usepackage{verbatim}

\normalsize
\newcommand{\req}[1]{(\ref{#1})}
\newcommand{\be}{\begin{equation}}
\newcommand{\ee}{\end{equation}}
\newcommand{\bea}{\begin{eqnarray}}
\newcommand{\eea}{\end{eqnarray}}

\newcommand{\BE}{\begin{eqnarray}}
\newcommand{\EE}{\end{eqnarray}}
\newcommand{\BEn}{\begin{eqnarray*}}
\newcommand{\EEn}{\end{eqnarray*}}
\newcommand{\barr}{\begin{array}}
\newcommand{\earr}{\end{array}}

\newcommand{\bit}{\begin{itemize}}
\newcommand{\eit}{\end{itemize}}
\newcommand{\bc}{\begin{center}}
\newcommand{\ec}{\end{center}}
\newcommand{\ben}{\begin{enumerate}}
\newcommand{\een}{\end{enumerate}}

\begin{document}

\title{Taking a shower in Youth Hostels: risks and delights of heterogeneity}
\author{Christina Matzke}
\affiliation{Bonn Graduate School of Economics, Department of Economics,
University of Bonn, Adenauerallee 24-26, D-53113 Bonn, Germany}
\email{christina.matzke@uni-bonn.de}
\author{Damien Challet}
\affiliation{Physics Department, University of Fribourg, P\'erolles, CH-1700 Fribourg, Switzerland
}
\email{damien.challet@unifr.ch}
\begin{abstract}
Tuning one's shower in some hotels may turn into a challenging
coordination game with imperfect information. The temperature
sensitivity increases with the number of agents, making the problem
possibly unlearnable. Because there is in practice a finite number of
possible tap positions, identical agents are unlikely to reach even
approximately their favorite water temperature. We show that a
population of agents with homogeneous strategies is evolutionary
unstable, which gives insights into the emergence of heterogeneity,
the latter being tempting but risky.
\end{abstract}
\keywords{coordination, heterogeneity, adaptive learning}
\maketitle

\section{Introduction}\label{introduction}

Taking a shower can turn into a painful tuning and retuning game when many people take a shower at the same time if the flux of hot
water is insufficient. In this fascinating game, it is in the interest of everybody not only to reach an agreeable equilibrium
temperature but also to avoid large fluctuations. These two goals are difficult to achieve because one inevitably not only has
incomplete information about the behavior and personal preferences of the other bathers, but also about the non-linear intricacies of
the plumbing system.

The central issue of this paper is to find the conditions under which the agents are satisfied, which depends on the learning
procedure and on its parameters.

The need to depart from rational representative agents was forcefully
voiced among others by Kirman \cite{kirman_06} and Arthur and Brian Arthur,
for instance in his El Farol bar problem \cite{Arthur}, subsequently simplified as Minority Game \cite{CZ97,MGbook}, from which we
shall borrow some ideas concerning the learning mechanism. In these models, the agents try to behave maximally differently from each
other, hence the need for heterogeneous agents.

The Shower Temperature Problem is different in that the perfect equilibrium is obtained when all the agents behave exactly in the
same optimal, unique way. A priori, it is a perfect example of a case where the representative agent approach applies fully. As we
shall see, however, because in practice there is a maximum number of tap tuning settings, it may pay off to be heterogeneous with
respect to the strategy sets. Therefore, the problem we propose in this paper is another example of a situation where heterogeneity
is tempting because potentially beneficial. The intrinsic and strong non-linearity of the temperature response function prevents the
use of the mathematical machinery for heterogeneous systems that successfully solved the Minority Game \cite{MGbook,CoolenBook}, the
El Farol bar problem \cite{CMO03} and the Clubbing problem \cite{TobiasClubbing}.

\section{The Shower Temperature Problem}\label{problem}

One of the problems of poor plumbing systems is the interaction between the water temperatures of all the people taking a shower
simultaneously. If one person changes her shower setting, she influences the temperature of all the other bathers. Cascading shower
tuning and retuning may follow. A key issue is how people can learn from past temperature fluctuations how to tune their own shower
so as to obtain an average agreeable temperature $\hat T$, and also to avoid large temperature fluctuations.

Some rudimentary shower systems allow only for one degree of freedom, the desired fraction of hot water in one's shower water,
denoted by $\phi \in [0,1]$. Assuming that $H$ and $C$ denote the maximal fluxes of hot and cold water available to a shower, and
that the total flux at this shower is constant, the obtained temperature is equal to \begin{equation} T=\frac{\phi H T_H+C
T_C(1-\phi)}{\phi H+C(1-\phi)},\end{equation} where $T_H$ and $T_C$ denote the constant temperatures of hot and cold water.

In the following, we shall consider the special case were $H=C$, $T_C=0$, and $T_H=1$, which amounts to express $T$ in $T_H$ units,
i.e., to rescale $T$ by $T_H$, which leads to $T=\phi$.

The situation may become more complex however if many people take a shower at the same time. Indeed, it sometimes happens that
altogether the $N$ bathers ask for a larger hot water flux than the plumbing system can provide, a feature more likely found in
old-style youth hostels than in more upmarket hotels (hence the title). Assume that the total available hot water flux for all
bathers together is $H$ while the cold water flux available at each single shower is $C=H$. We denote by $\Phi=\sum_{i=1}^N\phi_i$
the total fraction of asked hot water. If  $\Phi>1$, each agent will only receive $\phi_i/\Phi$ instead of $\phi_i$ and the total
flux of hot water she obtains is smaller than expected.\footnote[1]{The fraction of cold water in this case is still $1-\phi_i$,
according to the agent's choice, since cold water is assumed to be unrestricted.} Finally, agent $i$ obtains \begin{equation}
T_i=\frac{\phi_i}{\phi_i +\Psi(1-\phi_i)}, \label{eq:Ti}\end{equation}
 where $\Psi=\max(1,\Phi)$. Clearly, $T_i(\phi_i=0)=0$ and $T_i(\phi_i=1)=1$. When $\Phi\le1$, this equation reduces to the no-interaction case
 $T_i=\phi_i$. Therefore,
 provided that $\Phi>1$, the agents interact through the temperature they each obtain, that is, via $\Phi$. Assuming no inter-agent communication, the
 global quantity $\Phi$ is the only means of interaction. Therefore, this model is of mean-field nature. Henceforth, we consider
the more involved case of interaction, i.e., $\Phi>1$.

\section{Tuning one's shower}

\subsection{Equilibrium and sensitivity: the homogeneous case}\label{eq&sens}

Before setting up the full adaptive agent model, we shall discuss the homogeneous case where $\phi_i=\phi$.

Assuming that all the agents have the same favorite temperature ($\hat T_i=\hat T\le1$), they do not interact if $N\le1/\hat T$, in
which case $\phi=\hat T$. If $N>1/\hat T$ the equilibrium is reached
when 
\begin{equation}\label{eq:phi_eq} \phi=\phi_{\rm
eq}=1-\frac{1}{N}\left(\frac{1}{\hat T}-1\right). \end{equation}
 Hence, there is always a $\phi$ that satisfies everybody (for
instance, setting $\hat T=1/2$ leads to $\phi_{\rm eq}=1-1/N$). In equilibrium each agent actually gets
$\phi_{eq}H/(N\cdot\phi_{eq})=C/N$ hot water instead of $\phi_{eq}H$ and thus a total water flux of $C/N+(1-\phi_{eq})C=C/(N\hat T)$.
Hence, indeed the desired temperature $\hat T$ is reached for every agent, but the total water flux per agent is quite small for
large $N$.

\begin{figure}[t]
 \centerline{\includegraphics[width=0.6\textwidth]{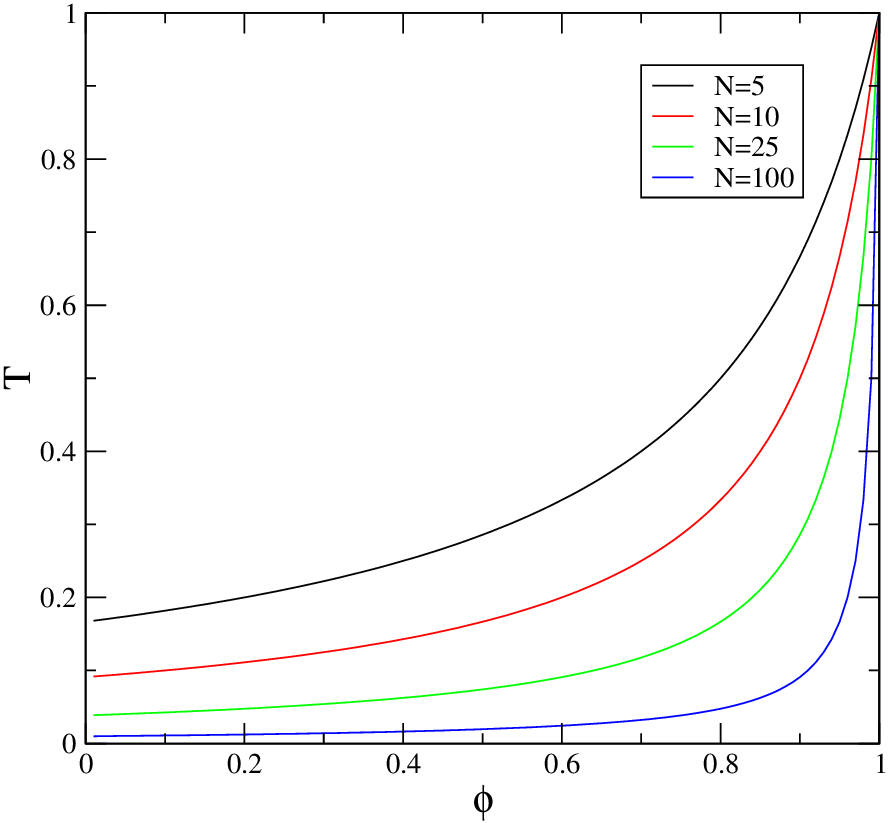}}
\caption{Individual temperature as a function of $\phi$ in the
homogeneous case for increasing $N$ (from top to bottom).}
\label{fig:T_vs_phi}
\end{figure}

The sensitivity of $T$ to $\phi$, defined as $\chi=\frac{\textrm{d} T}{\textrm{d} \phi}=\frac{N}{[1+N(1-\phi)]^2}$ is an increasing
function of $\phi$ and maximal at $\phi=1$ (a similar result also holds for $T_i=\frac{\phi_i}{\phi_i+\Phi(1-\phi_i)}$). The problem
is that $\chi(\phi_{\rm eq})= N\hat T^2\propto N$; therefore, as $N$ increases, tuning $\phi$ around $\phi_{\rm eq}$ becomes more and
more difficult, suggesting already that the agents might experience difficulties to learn how to tune their shower. Figure
\ref{fig:T_vs_phi} illustrates this phenomenon: as $N$ increases, the region in which most of the variation of $T$ occurs shrinks
substantially.

This problem is made worse by the fact that, in practice, there is only a finite number $S_{\rm max}$ of $\phi$s that can be
effectively used by the agents, mostly because of internal tap static friction---the larger the friction, the smaller the number of
different achievable $\phi$s. Assuming that the resolution in $\phi$ is $\delta \phi$, or equivalently that $S=1/(\delta\phi)$ values
of $\phi$ are usable, it becomes impossible to tune one's shower if $|T(\phi_{\rm eq}\pm\delta\phi)-\hat T|\simeq\chi(\phi_{\rm
eq})\delta \phi$ is larger than some acceptable value. As $\chi\propto N$ around $\phi_{\rm eq}$, $S\propto N$ is needed; as a
consequence, the ideal temperature is not learnable beyond a number of agents, which is for a large part pre-determined by the
plumbing system.

\subsection{Learning}\label{learning}

The question is how to reach $\phi_{\rm eq}$. In this model, it is hoped that the agents have a common interest to avoid large
fluctuations of $T_i$ around their favorite temperature $\hat T_i$: the Shower Temperature Problem is a repeated coordination game
(cf.\ \cite{crawford_haller_90} and \cite{bhaskar_00}) with many agents and limited information.

The dynamics of the agents are fully determined by their possible tap settings, thereafter called strategies, and by the trust they
have in them. Each agent $i$ has $S$ possible strategies $\phi_{i,s}$ with $s=1,...,S$ chosen in $[0,1]$ before the game begins and
kept constant afterwards (how to choose the $\phi$s is discussed in the next section). The typical resolution in $\phi$ is $1/S$; for
the same reason, the typical maximal $\phi_i$ over all the agents is of order $1-1/S$. This paper follows the road of inductive
behavior advocated by Brian Arthur: to each possible choice $\phi_{i,s}$ agent $i$ attributes a score $U_{i,s}(t)$ (where $t$ denotes
the time step of the game), which describes its cumulated payoff at time $t$. The agents choose probabilistically their $\phi_{i,s}$
according to a logit model $P(\phi_i(t)=\phi_{i,s})=\exp(\Gamma U_{i,s}(t))/Z$, where $Z$ is a normalization factor and $\Gamma$ is
the rate of reaction to a relative change of $U_{i,s}$.

If one were to follow blindly El Farol bar problem and Minority Game literature, one would write
$$
U_{i,s}(t+1)=U_{i,s}(t)+\phi_{i,s}\left[\hat T_i-T_i(t)\right].
$$
When $S>2$, such payoffs are not suitable any more, as the agents switch between their highest and smallest $\phi_{i,s}$, the
intermediate ones being sometimes used only because of fluctuations induced by the stochastic strategy choice. A payoff allowing for
a gradual increase of $\phi_{i,s}$ is necessary. Absolute value-based payoffs are fit for this purpose\footnote{Quadratic payoffs,
albeit mathematically sound, are more problematic for performing numerical simulations.}: mathematically,
$$
U_{i,s}(t+1)=U_{i,s}(t)-\left|\hat T_i-T_i(t)\right|.
$$
This payoff however does not depend on $\phi_{i,s}$. As a consequence, all the strategies have the same payoff. Therefore, one has to
give more information to the agents. An agent that has perfect information about the plumbing system, the temperatures and fluxes of
hot and cold water --- for instance the plumber that built the whole installation --- may know precisely which temperature she would
have obtained, had she played $\phi_{i,s'}$ instead of her chosen action $\phi_{i,s_i(t)}$. Such people are probably not very
frequent amongst the general population, however. This is why we shall consider an in-between case, where the agents' estimation of
$T_{i,s}(t)$ is a linear interpolation between the temperature of the strategy currently in use, i.\,e.\ $T_i(t)=T_{i,s_i(t)}$ and
its correct virtual value. The payoff is therefore 
\begin{equation} U_{i,s}(t+1)=U_{i,s}(t)(1-\lambda)-\lambda\left|\hat
T_i-(1-\eta)T_i(t)-\eta T_{i,s}(t)\right|,\label{eq:payoff}
\end{equation}
 where $\eta\in [0,1]$ encodes the ability of the agents to
infer the influence of $\phi_{i,s}$ on the real temperature and $0\le\lambda<1$ introduces an exponential decay of cumulated payoffs,
with typical score memory length $\propto 1/\lambda$. The parameter $\eta$ is related to the difference between naive and
sophisticated agents as defined by Rustichini \cite{rustichini}. The first kind of agents believe that they are faced with an external process,
i.\,e.\ that they do not contribute to $\Phi$, whereas sophisticated agents are able to compute $\Phi_{-i}=\Phi-\phi_i$. In this
model, perfect sophisticated agents have $\eta=1$.

\section{Results}

It is natural to measure two collective quantities, the average temperature $T$ obtained by the agents and its average distance from
ideal temperature averaged over all the agents, denoted by ${\Delta T}=T-\hat T$; this characterizes the average temperature obtained
by the agents, or how far the agents are collectively from their goal. The individual dissatisfaction is the distance from the ideal
temperature for a given agent; one therefore measures it with ${|\delta T|}=\frac1N\sum_{i=1}^N|T_i-\hat T_i|$; it is a measure of
the average risk.

All the quantities reported here are measured in the stationary state over $10,000$ time steps for $\hat T=0.5$, $\eta=1$,
$\lambda=0.001$ and if not stated differently $N=20$, after an equilibration time of $30/(\lambda \Gamma)$. The stationary state does
not depend much on $\lambda$. On the other hand, the performance of the population is of course improved as $\eta$ increases and
saturates for $\eta>0.5$. The role of $\Gamma$ is discussed below.

\subsection{Homogeneous population}

Since the equilibrium is reached when all the agents tune their shower in exactly the same way, trying first homogenous agents (or
equivalently a representative agent) makes sense {\em a priori}. We shall therefore set $\phi_{i,s}=\phi_{s}=\frac{s}{S+1}$,
$s=1,...,S$ so that the agents avoid using only hot or cold water.

\begin{figure}[t]
 \centerline{\includegraphics[width=0.6\textwidth]{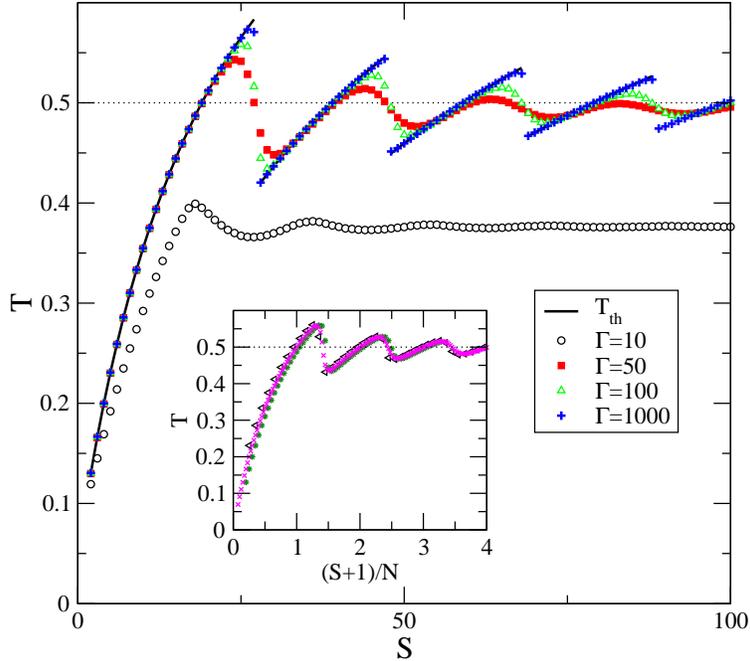}}
\caption{Temperature $T$ reached by homogeneous agents as a function
of $S$ for various $\Gamma$. Inset: $T$ vs. $(S+1)/N$, showing the
scaling property of $T$, with $N=10,20,40$ (asterisks, triangles,
crosses).} \label{fig:T_vs_S}
\end{figure}

Agents with homogeneous strategies have a peculiar way of converging to their ideal temperature as $S$ increases. Figure
\ref{fig:T_vs_S} displays the oscillations of the reached temperature with decreasing amplitude as a function of $S$. The asymmetric
upward and downward slopes are due to the asymmetry of $T$ around $\phi_{\rm eq}$, as seen in Fig. \ref{fig:T_vs_phi}.
Theoretically, this can easily be explained by assuming that all the agents select the same $s$ that gives $T$ as close as possible
to $\hat T$. If $s$ was a real number, $\hat s=[1-1/N(1/\hat T-1)](S+1)$. The choice of the agents therefore is limited to $[\hat s]$
and $[\hat s]+1$ where $[x]$ is the integer part of $x$ (one may need to enforce $[\hat s]<S$ when $S<N$). $T([\hat s])$ and $T([\hat
s]+1)$ are alternatively closest to $\hat T$, therefore this actual optimal temperature $T_{\rm th}$ (whichever $T([\hat s])$ or
$T([\hat s]+1)$) oscillates around $\hat T$, as seen in Fig. \ref{fig:T_vs_S}. The period of the oscillations is $N$, and their
amplitude decreases as $1/S$. As expected, a very large value of $\Gamma$ replicates closely the dented nature of the value of
$T_{\rm th}$, in which case all the agents take the same choice even close to the peak of $T_{\rm th}$. Generally, smaller $\Gamma$s
(at least to a certain degree) lead to better average temperatures as it allows to play mixed strategies, and thus combine two
temperature so as to achieve a collective average result closest to $\hat T$. From that point of view, $\Gamma=50$ is a better choice
than $\Gamma=1000$. Hence, there exists an optimal global value of $\Gamma$, leading to a mixed-strategy equilibrium. This is because
taking stochastic decisions is a way to overcome the rigid structure imposed on the strategy space, whose inadequacy is reinforced by
the strong non-linearity of $T(\phi)$. Too small a $\Gamma$ is detrimental as it allows for using $\phi$ further away from $\phi_{\rm
eq}$; because of the shape of $T(\phi)$, those with smaller $\phi$ are more likely to be selected.

\begin{figure}[t]
 \centerline{\includegraphics[width=0.6\textwidth]{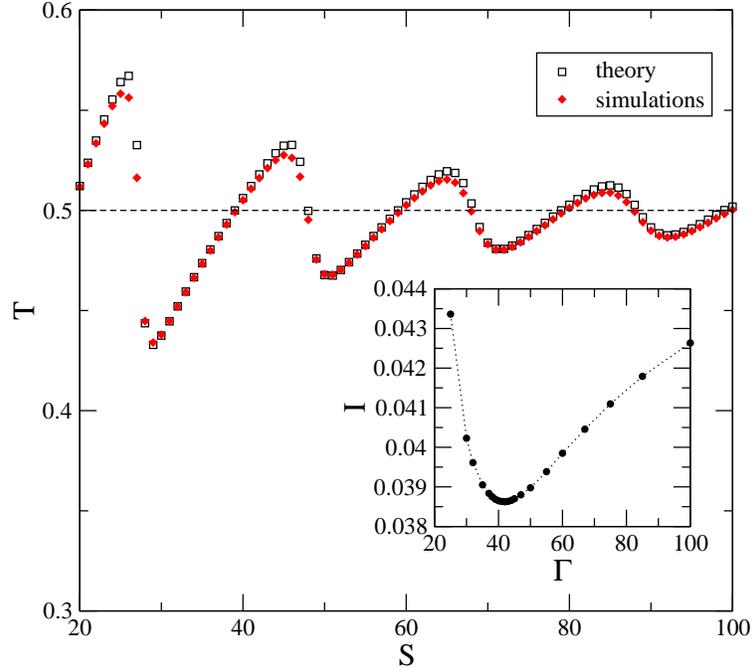}}
\caption{Temperature $T$ reached by homogeneous agents as a function
of $S$ for $\Gamma=100$. Squares: theory, circles: numerical
simulations. Inset: average deviation $I$ from $\hat T$ versus
$\Gamma$ (same parameters); the dotted lines are for eye guidance
only.} \label{fig:Tth_vs_S}
\end{figure}

The individual dissatisfaction $|\delta T|$ unsurprisingly mirrors $|\Delta T|$ since all the players are identical. Both quantities
are the same for large $\Gamma$ as everybody plays the same fixed strategy. $|\delta T|$ also decreases as $1/S$ (see Fig.
\ref{fig:dT_vs_S}). However, the larger $\Gamma$, the smaller $|\delta T|$, as each agent manages to get closer to the optimal
choice.

It is easy to obtain analytical insights by solving the stationary
state equations for $U_{i,s}$ \req{eq:payoff}. For the sake of simplicity,
assuming that $\eta=1$ and that only the two $\phi$s surrounding $\phi_{\rm eq}$, i.\,e.\ $[\hat s]$ and $[\hat s]+1$, denoted by $-$
and $+$ respectively, are used, one obtains the set of equations (independent from $\lambda$ and $i$) \begin{equation}
U_{i,\pm}=U_\pm=-|T_\pm-\hat T| \end{equation} where
\begin{equation}
T_{i,\pm}=T_\pm=\frac{1}{1+\frac{N_+\phi_++N_-\phi_-}{\phi_\pm}(1-\phi_\pm)}
\end{equation} with $N_\pm=N\cdot P(s=\pm)$, where
$P(s=+)=\frac{\exp(\Gamma U_{i,+})}{\exp(\Gamma U_{i,+})+\exp(\Gamma U_{i,-})}$ and $P(s=-)=1-P(s=+)$ is a Logit model for the
two-strategy case $S=2$. Figure \ref{fig:Tth_vs_S} shows the good agreement between numerical simulations and this simple theory,
especially in the convex part of the oscillations, as long as $\Gamma$
is large enough to prevent the use of more than 2
strategies.

Being faced with oscillations (as a function of $S$ or $N$) of the
expected value of $T$ is problematic for homogeneous agents since they do not know $N$
a priori and because $N$ may vary with time, leading to dramatic
shifts of $\hat T$. In addition, since all
the agents select the same $\phi$ for large $\Gamma$, not a single
agent is ever likely to reach a temperature close to $\hat T$. The
agents do not know whether on average they will overheat or chill. A
way to measure this uncertainty is to measure the average $|{\Delta
  T}|$ over $S$ in numerical simulations, for instance with
$I=\sum_{S=N}^{5N}|{\Delta T}|/(4N)$.\footnote{Simulations show that
  the average temperature is in fact a function of $(S+1)/N$
  (cf.\ Fig.\ \ref{fig:T_vs_S}) (instead of a function of $S$ and
  $N$), i.\,e.\ Fig\ \ref{fig:Tth_vs_S} would look the same if $S$
  was fixed and $N$ varied. Hence we may take the average over $S$
  instead of over $N$.} The inset of Fig\ \ref{fig:Tth_vs_S} reports
that the minimum of $I$ is at $\Gamma\simeq 42$ for the chosen
parameters, which shows the existence of an optimal learning
rate. Since the individual satisfaction is maximal in the limit
$\Gamma\to\infty$ (see above) there is no minimum of a similar measure
for $|\delta T|$.

\subsection{Heterogeneous populations}
\begin{figure}[t]
 \centerline{\includegraphics[width=0.6\textwidth]{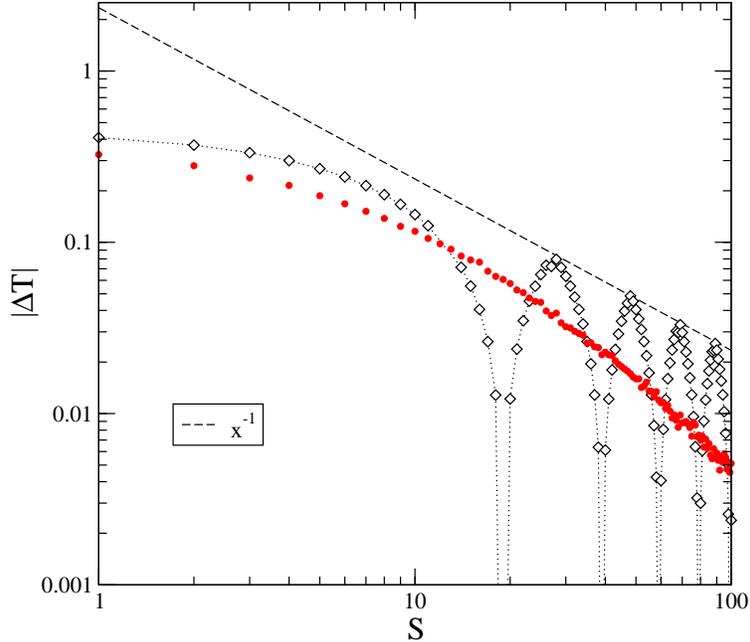}}
\caption{Absolute temperature deviation $|{\Delta T}|$ reached by homogeneous (squares) and heterogeneous (circles) agents
 as a function of
$S$ for $\Gamma=100$. Average over 500 samples for heterogeneous
agents.} \label{fig:absdT_vs_S}
\end{figure}

There are many ways for agents to be heterogeneous. One could imagine to vary $S$, $\Gamma$, $\eta$, $\lambda$ or $\hat T$ amongst
the agents. Here we focus on strategy heterogeneity, i.\,e.\ the agents face showers with different tap settings: the strategy space
of agent $i$ is no longer $\frac{1}{S+1},\ldots,\frac{S}{S+1}$, but now each agent has an individual strategy space where each
strategy $\phi_{i,s}$, $s=1,\ldots,S$, is assigned a random number from the uniform distribution on $[0,1]$ before the simulation.

Intuitively, the effect of heterogeneity is to break the structural rigidity of the strategy set of a representative agent. Figure
\ref{fig:absdT_vs_S} reports that $|{\Delta T}|$ does not oscillate, but converge (from below) faster than $S^{-1}$ to zero.
Homogeneous agents might achieve a better average temperature depending on $N$ and $S$,
but on the whole clearly perform collectively worse. This is simply
because most likely homogeneous agents have a $\phi$ whose difference
with $\phi_{eq}$ is smaller than $1/(S+1)$ 
In addition, heterogeneous agents expect to have a smaller than ideal temperature, but on average {\em predictably smaller}, with no
strong dependence on $S$. Thus, the expectation over the temperature of the agents is much improved by heterogeneity.

\begin{figure}[t]
 \centerline{\includegraphics[width=0.8\textwidth]{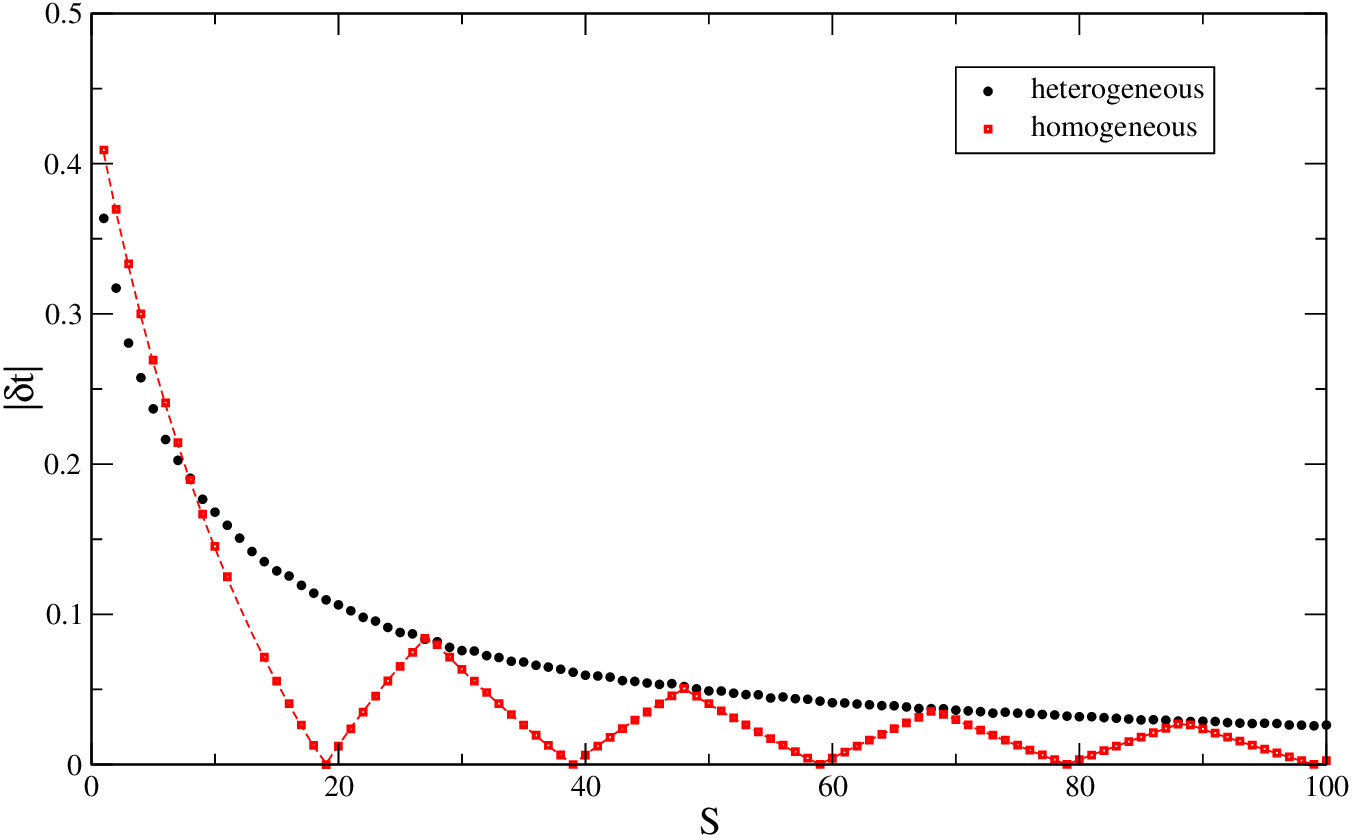}}
\caption{Individual dissatisfaction $|{\delta T}|$ reached by homogeneous (empty squares) and heterogeneous agents (full circles)
 as a function of
$S$ for $\Gamma=1000$. Average over 500 samples for heterogeneous
agents. Dashed line: theoretical predictions.} \label{fig:dT_vs_S}
\end{figure}

However, looking at the average absolute individual deviation from $\hat T$ reveals that the uncertainty brought by heterogeneity is
considerably worse {\em on average}. Plotting $|\delta T|$ for both types of agents shows that $|\delta T|$ is always smaller for
homogeneous agents (Fig. \ref{fig:dT_vs_S}). This means that if being heterogeneous is more risky. Which agent (or equivalently,
shower) performs better depends not only on $N$, but also on the tuning settings of all the agents.

\subsection{Homegeneous vs heterogeneous}

\begin{figure}[t]
 \centerline{\includegraphics[width=0.6\textwidth]{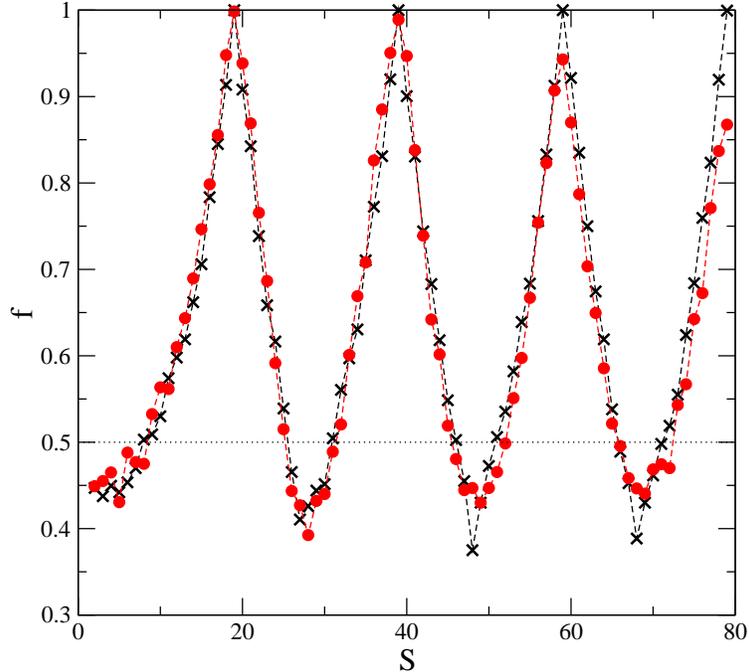}}
\caption{Fraction of the runs for which a single heterogeneous agent
is worse off than the other $N-1$ homogeneous agents; $\Gamma=1000$
(crosses) and $\Gamma=30$ (circles). Average over 2000 samples.}
\label{fig:f_vs_S}
\end{figure}

Heterogeneity may be tempting as it suppresses the systematic abrupt
oscillations experienced by homogeneous populations when 
$N$ changes and is
collectively better on average. However,  heterogeneous showers are potentially more risky. In other words, the agents
must consider the trade-off between the temptation of an expected better temperature and a potentially larger deviation.

The situation discussed above is only global. Does it pay to be heterogeneous for a single agent? An answer comes from  a system
consisting of $N-1$ homogeneous agents as defined above and a single random one with random $\phi_{i,s}$s. The fraction $f$ of the
runs at fixed $S$ that give a better $\delta T_i$ to the homogeneous showers is reported in Fig. \ref{fig:f_vs_S}; this quantity
indicates that the majority of heterogeneous agents are not worse off for about a quarter of the values of $S$. This finding is not
in contradiction with the fact that the average personal dissatisfaction of heterogeneous agents is always larger than that of
homogeneous agents: $|\delta T|$ is much influenced by large deviations contributed by  a minority of agents because of large
temperature sensitivity to small deviations in $\phi$. Finally, the advantage of the homogeneous population increases with $\Gamma$,
as a large learning rate helps only using one's best strategy.

\begin{figure}[t]
 \centerline{\includegraphics[width=0.6\textwidth]{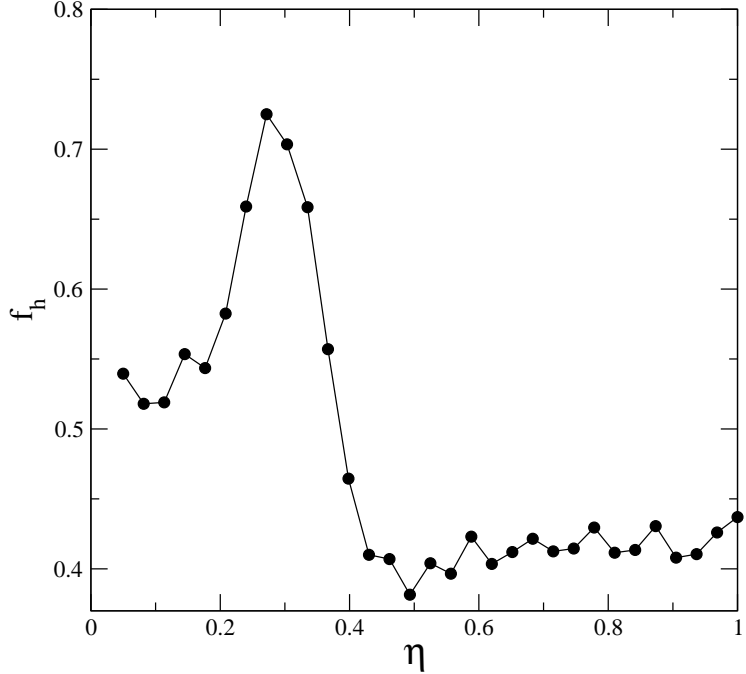}}
\caption{Fraction of the agents using one of their homogeneous
  strategies as a function of $\eta$; $\Gamma=100$. Average over 2000 samples.}
\label{fig:f_vs_eta}
\end{figure}

Let us finally give to all agents the possibility to use either
strategies from the homogeneous set, or strategies drawn at random. A
simple way to achieve this is to give the agents $2S$ strategies, $S$
of them defined homogeneously, and $S$ of them drawn at random before
the game begins. We shall then be interested in
$f_\mathrm{h}$, the average fraction of players using
strategies from the homogeneous set. It turns out that when $\eta=1$,
this fraction fluctuations as a function of $S$, for instance, but
remains roughly constant. A more interesting behaviour come from
varying $\eta$ (see Fig. \ref{fig:f_vs_eta}). When $\eta=0$, the
population is not expected to show any preference since all the
score updates are the same for a given agent. Then, as $\eta$ is
increased, the discrimination power of the agents improves. Quite
peculiarly, a peak of advantageous homongeneity arises around
$\eta\simeq 0.3$. The saturation of $f_\mathrm{h}\simeq0.42$ for $\eta>0.5$ shows
that in that case most agents stick to a heteregeneous
strategy. Still, homogeneity and heterogeneity coexist. This probably
comes from the statistical properties of distributions of random
strategies around $\phi_{eq}$ together with the very strong
non-linearity of $\hat T$ as a function of $\phi$

\section{Discussion and conclusions}\label{conclusion}

As a final note, minimizing $|\Delta T|$ is equivalent to solving a number partitioning problem \cite{GareyJohnson} in which one
splits a set of $N$ numbers $a_i>0$ into two subsets, so that the sums of the numbers in the subsets are as close as possible, which
amounts to minimize $C=|\sum_i s_i a_i|$ where $s_i=\pm1$; it is an NP-complete problem; in other words, the only way to find the
absolute minimum of $C$ is to sample all the $2^N$ configurations. Let us consider an even simpler version of the Shower Temperature
Problem that makes more explicit its NP-complete nature. Each agent $i$ is given $a_i$ and plays $\phi_{\rm eq}+s_i a_i$, $s_i=\pm1$.
Neglecting the self-impact on the resulting temperature and the non-linearity of the temperature response, the analogy between the
Shower Temperature Problem and the number partitioning problem is straightforward. Methods borrowed from statistical mechanics show
that the average optimal $C$ scales as $2^{-N}$, which requires to enumerate  the $2^N$ possible configurations \cite{MertensPRL1}.
This is much better than what the agents achieve; the reason for this discrepancy is that the agents do not reach a stationary state
in $O(\exp N)$ time steps, hence, they cannot sample all the possible configurations. Another reason is that the optimal solution may
require some agents to use a strategy that would yield a worse temperature than their optimal choice.

In conclusion, the Shower Temperature Problem shows the subtle trade-offs between a homogeneous population with equally spaced
actions and a fully random one. In a system where the agents' action space is not likely to include the optimal equilibrium choice,
heterogeneity is a way to solve more robustly, with less systematic
deviation on a collective level this kind of problem, at the expense
of a higher risk for individual agents. In other words,  if given
the choice, some agents favoured by randomness will take the
opportunity to improve their fate. Therefore even simple situations
where a simple representative agent approach yields a unique optimal
choice, it may not be reachable because of practical
constraints. And thus heterogeneity emerges.

\bibliography{thebibliography}







\end{document}